%% file: ms.tex
\documentclass{article}

\usepackage{arxiv}

\usepackage[numbers]{natbib}
\usepackage{graphicx}
\usepackage{amsmath,amsfonts}
\usepackage{times}
\usepackage{t1enc}
\usepackage{graphicx}
\usepackage{textcomp}
\usepackage{color,xcolor}
\usepackage{epstopdf}
\usepackage{rotating}
\usepackage{xifthen}
\usepackage{paralist}
\usepackage{algorithm}
\usepackage{soul}
\usepackage[nointegrals]{wasysym}
\usepackage{algpseudocode}
\usepackage{multirow}
\usepackage{amsthm}
\usepackage[none]{hyphenat}
\theoremstyle{plain}
\newtheorem{theorem}{Theorem}
\newtheorem{definition}{Definition}
\newtheorem{lemma}         [theorem]{Lemma}

\newtheorem*{remark} {Remark}

\usepackage{nameref}
\usepackage{varioref}
\usepackage{array}
\newcolumntype{P}[1]{>{\centering\arraybackslash}p{#1}}
\newcolumntype{M}[1]{>{\centering\arraybackslash}m{#1}}
\usepackage{cleveref}
\algrenewcommand\algorithmicrequire{\textbf{Input:}}
\algrenewcommand\algorithmicensure{\textbf{Output:}}

\newcommand{\smite}{\emph{smite-node}}
\newcommand{\fault}{\emph{faulty-node}}
\usepackage{comment}
\usepackage{array}

\usepackage[symbol]{footmisc}

\usepackage{amsmath, amsthm, enumerate,boxedminipage}
\usepackage{xcolor}
\colorlet{RED}{red}
\colorlet{BLUE}{blue}
\colorlet{TEAL}{teal}

\newboolean{short}
\setboolean{short}{false}

\newcommand{\shortOnly}[1]{\ifthenelse{\boolean{short}}{#1}{}}
\newcommand{\onlyShort}[1]{\ifthenelse{\boolean{short}}{#1}{}}
\newcommand{\longOnly}[1]{\ifthenelse{\boolean{short}}{}{#1}}
\newcommand{\onlyLong}[1]{\ifthenelse{\boolean{short}}{}{#1}}
\newcommand{\shortLong}[2]{\ifthenelse{\boolean{short}}{#2}{#1}}
\newcommand{\longShort}[2]{\ifthenelse{\boolean{short}}{#2}{#1}} 

\AtBeginDocument{%
  }
\title{Fault-Tolerant Graph Realizations in the Congested Clique}

\begin{document}

\makeatletter
\newcommand{\linebreakand}{%
  \end{@IEEEauthorhalign}
  \hfill\mbo

  \mbox{}\hfill\begin{@IEEEauthorhalign}
}
\makeatother
\author{Anisur Rahaman Molla, Manish Kumar, Sumathi Sivasubramaniam\\
Indian Statistical Institute Kolkata}

\maketitle

\begin{abstract}
  In this paper, we study the graph realization problem in the Congested Clique model of distributed computing under crash faults. We consider {\em degree-sequence realization}, in which each node $v$ is associated with a degree value $d(v)$, and the resulting degree sequence is realizable if it is possible to construct an overlay network with the given degrees. 
  Our main result is a $O(f)$-round deterministic algorithm for the degree-sequence realization problem in a $n$-node Congested Clique, of which $f$ nodes could be faulty ($f<n$). The algorithm uses $O(n^2)$ messages. We complement the result with lower bounds to show that the algorithm is tight w.r.t the number of rounds and the messages simultaneously.
  We also extend our result to the Node Capacitated Clique (NCC) model, where each node is restricted to sending and receiving at-most $O(\log n)$ messages per round. In the NCC model, our algorithm solves  degree-sequence realization in $O(nf/\log n)$ rounds and $O(n^2)$ messages. 
  For both settings, our algorithms work without the knowledge of $f$, the number of faults. To the best of our knowledge, these are the first results for the graph realization problem in the crash-fault distributed network.
  \vspace{0.01cm}
\end{abstract}

\input{introduction}

\input{related_work}

\input{without_knowing_f}

\input{ncc-extension}

\input{conclusion}

\vspace{1 cm}



\end{document}

%% file: introduction.tex
\section{Introduction}\label{sec: introduction}
    {\em Graph Realization} problems have been studied extensively in the literature, mainly in the sequential setting. In general, graph realization problems deal with constructing graphs that satisfy certain predefined properties (such as a degree sequence). While the area was mostly focused on realizing graphs with specified degrees \cite{H55},  other properties, such as connectivity \cite{FC70,F92, F94}, flow \cite{GH61} and eccentricities \cite{B76, L75} have also been studied.
    
     The \emph{degree-sequence} realization problem has been explored widely in the centralized setting. Typically, the problem consists of the following. Given a  
     sequence of non-negative numbers $D = (d_1, d_2, \ldots, d_n)$, the degree-sequence problem asks if $D$ is \emph{realizable}. A sequence $D$ is said to be {\em realizable} if there is a graph of $n$ nodes whose sequence of degrees matches $D$. The first complete characterization of the problem was established in 1960, when Erd\"os and Gallai~\cite{EG60}  showed that $D$ is realizable if and only if $\sum_{i=1}^k d_i \le k(k-1) + \sum_{i=k+1}^n \min(d_i, k)$ for every $k \in [1,n]$ 
     following which, the definitive solution was independently found by  Havel and Hakimi~\cite{H55,Ha62}, a recursive sequential algorithm that takes $O(n)$ time (see Section~\ref{sec:shh} for more detail). Non-centralized versions of realizing degree sequences have also been studied, albeit to a lesser extent.  Arikati and Maheshwari~\cite{AM1996realizing} provide an efficient technique to realize degree sequences in the PRAM model. 
     
    
    Recently, Augustine et al. \cite{ACCPSS22} studied the graph realization problem in distributed networks. They approach the graph realization from the perspective of Peer-to-Peer (P2P) overlay construction. P2P overlay networks are virtual networks built on top of the underlying network, e.g., Internet. Given the increasing popularity of P2P networks (in the key areas like, cryptocurrencies, blockchain, etc.), research on P2P networks, particularly in the construction of P2P overlays, has become a crucial part in distributed computing.
    
    In \cite{ACCPSS22}, the authors build overlay networks by adapting graph realization algorithms. Briefly, given a  network of $n$ nodes $V=\{v_1,\ldots,v_n\}$, where each $v_i$ is assigned a degree $d_i$,  the goal is to create an overlay graph $G(V,E)$ such that $d(v_i)=d_i$ in $G$, and for any edge $e\in E$, at least one of $e$'s end points knows the existence of $e$ (referred to as an \emph{implicit realization}).  Note that, for any edge it is only required that one end point must be aware of its existence, which means that a node may be aware of only a small part of the realized graph. In the best case, the nodes know only their neighbors in the final realization. However, in most of the P2P overlay applications, it is crucial to know the entire overlay graph. Furthermore, the network is assumed to be fault-free.
    
    In this work, we extend the graph realization problem in faulty setting where node may fail by crashing. In addition, we consider the graph realization from the approach of learning degrees so that the nodes may recreate the degree sequence locally, i.e., they know the entire overlay graph. To be precise, we develop fault-tolerant algorithm which ensures that each node learns the degree of all the non-crashed nodes (at least) in the network. This allows the nodes to use the sequential Havel-Hakimi algorithm ~\cite{H55,Ha62} to build the overlay graph $G$ locally. Note that, since every (non-crashed) node learns the degree sequence and creates the realization, all of them know the entire overlay graph. 
    
    Our work is primarily in the Congested Clique (CC) model, a well studied model in distributed networks~\cite{LSPP05}. The Congested Clique model has been explored widely for many fundamental distributed problems \cite{HDK21, GP16, GN18, J018, JN18, KS18, BMRT18, DKO14,konrad2018mis, BK18}. Distributed networks are faulty by nature; nodes crash, links fail, nodes may behave maliciously, etc. In the node failure settings, the two fundamental problems, namely, agreement and leader election has been studied extensively~\cite{DS82, GMY95, KKKSS08, AMP18, BPV06, FM97, KS16, ACDNP0S19, KM21, GK10}. The main challenge in a faulty setting is to ensure that no two nodes have a different view of a value, which may lead to an increase in the number of rounds or messages or both. Thus, in our work, we focus on developing an algorithm that minimizes the time complexity and the message complexity simultaneously in the Congested Clique with faulty nodes.

    In order to make our algorithm more suited for the networks with message overhead limitation (such as peer-to-peer networks), we also extend the algorithm to the Node Capacitated Clique (NCC) model, introduced by Augustine et al. \cite{AGGHSKL19}. In the NCC, nodes are restricted to send and receive only $O(\log n)$ messages per round, making it ideal for more realistic settings (e.g., less or no overhead at nodes). The authors in~\cite{ACCPSS22} restrict their graph realization algorithms primarily to the NCC model. While our main algorithm considers the CC model, we extend it to the NCC model also. For both the settings (CC and NCC), our algorithms do not require any prior knowledge of $f$--- the number of faulty nodes.
    
    As a byproduct, our algorithms solve fault-tolerant reliable broadcast~\cite{raynal-book18,rocsu1996early,rosu1996optimal} in a congested clique. Reliable broadcast algorithms ensure that both senders and receivers agree on a value $m$ sent by a correct process. That is, either all correct processes decide on the value $m$ being sent or decide the sender is faulty.  While this is similar to our problem, in our case we extend the basic premise to ensuring that all nodes agree on the same degree sequence, which consists of ensuring that the network agrees on at most $n$ values (non-faulty nodes' degree). This causes a significant challenge to the broadcast protocols in a faulty setting. To the best of our knowledge, the best known for reliable broadcast protocol in the crash failure setting has $O(f)$ time complexity and $O(n^2)$ messages message complexity~\cite{raynal-book18,rosu1996optimal}. Moreover, in a reliable broadcast one message is counted for one broadcast, while in our algorithm it is counted as $n$ messages in the congested clique.  Thus, the best known reliable broadcast protocol won't give better bounds if applies to solve the graph realization problem.

    \medskip
    
    \noindent \textbf{Paper Organization:} The rest of the paper is organized as follows. In the rest of the section, we first present our results, followed by a brief description of the model and a formal definition of our problem. In Section \ref{sec: related_work}, we discuss various related works in the field. Section \ref{sec:shh} provides a brief description of the sequential Havel-Hakimi solution for the graph realization problem. The Section \ref{sec: without knowing f} presents the main result, which is an efficient solution for the fault-tolerant graph realization problem and also provides matching lower bounds. In Section \ref{sec: NCC model}, we extend the graph realization algorithm to the NCC model. And finally, in Section \ref{sec: conclusion} we conclude with some interesting open problems.
    
    \subsection{Our Contributions}\label{subsec: contribution}
    We present an efficient algorithm and matching lower bounds for the distributed graph realization problem in the Congested Clique and Node Capacitated Clique model in the presence of node failures. Specifically, our results are: 
    
    \begin{enumerate}[(1)]
    	\item An $O(f)$-round deterministic algorithm for the graph realization problem in an $n$-node Congested Clique with at most $f < n$ node failures. The message complexity of the algorithm is $O(n^2)$, where the size of each message is $O(\log n)$ bits. 
    	
    	\item  We show a matching lower bound for both the time and the message bounds-- which demonstrates the simultaneous time and message bounds optimality of our algorithm.   
    	
    	\item We extend the algorithm for the Congested Clique to the Capacitated Clique, and present a $O(nf/\log n)$-round and $O(n^2)$-message complexity algorithm.  
    \end{enumerate}
    
    \subsection{Model and Definitions}\label{subsec:  model}
    We consider the message passing model of distributed computing. The underlying network is a Congested Clique (CC) \cite{KS18, LSPP05}. A Congested Clique consists of $n$ nodes which are identified by unique IDs from $\{u_1, u_2, \dots, u_n\}$. Communication among the nodes happens in synchronous rounds, where in each round a node can communicate with any of the other $n-1$ neighbors by sending a message of size $O(\log n)$ bits. This is known as the CONGEST communication model \cite{Pelege2000}.  Nodes know the ID of the other nodes and the port connecting to the IDs. This assumption is known as the $KT_1$ model \cite{Pelege2000}. We assume that an arbitrary subset of the nodes in the clique of size up to $f < n$ may fail by crashing. 
     A faulty node may crash in any round during the execution of the algorithm. If a node crashes in a round, then an arbitrary subset (possibly all) of its messages for that round may be lost (as determined by an adversary) and may not have reached the destination (i.e., neighbors). The crashed node halts (inactive) in the further rounds of the algorithm. If a node does not crash in a round, then all the messages that it sends in that round are delivered to the destination. 
     
     We assume an {\em adaptive} adversary controls the faulty nodes, which selects the faulty nodes at any time during the execution of the algorithm. Also, the adversary chooses when and how a node crashes, which is unknown to the nodes.   
    
    The time complexity of an algorithm is the number of rounds from the start until the termination of the algorithm. The message complexity of an algorithm is the total number of messages sent by all the nodes throughout the execution of the algorithm.
    
    In the interest of being useful in the Peer-to-Peer context, we also consider the Node Capacitated Clique (NCC) \cite{AGGHSKL19} model. NCC limits each node to send or receive a bounded number of messages, which, interestingly, makes NCC quite distinct from CC.  In this model, any node, say, $u$ can send or receive   $O(\log n)$ messages, each of size $O(\log n)$ bits.
    
    

    We will now formally define the  distributed graph realization problem (with and without faults). We say that an overlay graph $G=(V,E)$ is constructed if, for every $e = (u,v) \in E$, at least one of the endpoints is aware of the ID of the other and also aware that $e \in E$. We say that the overlay  graph is {\em explicit} if, for every edge $e \in E$ in the graph both endpoints know each other's ID and are aware of that $e \in E$. 
    
    \begin{definition}[Distributed Graph Realization \cite{ACCPSS22}]
    Let $V=\{v_1,\ldots, v_n\}$ be the set of nodes in the network. Let $D=(d_1,d_2,\ldots, d_n)$ be an input degree sequence such that each $d_i$ is only known to the corresponding node $v_i$. The distributed degree realization problem requires that the nodes in $V$ construct a graph realization of $D$ such that in the resulting overlay graph $G$, the following conditions hold:
    \begin{enumerate}[(i)]
    	\item The degree sequence of $G$ is precisely $D$.
    	\item The degree of $v_i$ is $d_i$, $\forall i \in \{1, \dots, n\}$.
    \end{enumerate}
    Thus, in the case of the distributed graph realization problem, a solution should output the graph $G$ if $D$ is a realizable degree sequence; otherwise, output “unrealizable”. 
    \end{definition}
    

    \begin{definition} [Distributed Graph Realization with Faults]\label{def : with fault}
    Let $V=\{v_1,\ldots v_n\}$ be the set of nodes in the network and $D=(d_1,d_2,\ldots, d_n)$ be an input degree sequence such that each $d_i$ is only known to the corresponding node $v_i$. Let $F\subset V$ be an arbitrary subset of faulty nodes in the network, such that $|F|=f \leq n-1$. 
    Let us define $D' \subseteq D$ be the modified degree sequence after losing the degrees of some faulty nodes 
    and $G'$ be the corresponding overlay graph over $D'$. The distributed degree realization with faults problem requires that the {\em non-faulty} nodes in $V$ construct a graph realization of $D'$ such that in the resulting overlay graph $G'$, the following conditions hold: 
    \begin{enumerate}[(i)]\label{def:grwithfaults}
        \item $|D'|\geq n-|F|$. 
        \item  $D-D'$ is the degree sequence corresponding to some nodes that crashed.
        \item For any edge $e = (u,v)\in G'$, either $u$ or $v$ (or both) knows of the existence of $e$.
    \end{enumerate}
    The required output is an overlay graph $G'$ if $D'$ is realizable; otherwise, output “unrealizable”.
    \end{definition}

%% file: related_work.tex
\section{Related Work}\label{sec: related_work}
Fault tolerant computation has always been a popular area of research in distributed computation, only becoming more popular with the prevalence of P2P networks that encourage for high decentralization. Often the focus of such research is on maintaining connectivity~\cite{Bagchi06}, recovery, or on ensuring that the network can tolerate a certain number of faults~\cite{U92}. We refer interested readers to~\cite{scheideler2002models} for more information on models and techniques for designing such systems. In our work, we focus mainly on ensuring that all the (non-faulty) nodes have the same view of the information in-spite of the presence of numerous faults.

In lieu of this, our goal in this work is to solve the distributed degree sequence problem, which is a graph realization problem. Graph realization problems have been well studied in the literature, focusing on problems such as realizing graphs with specified degrees~\cite{H55} as well as other properties, like connectivity and flow \cite{GH61,FC70,F92,F94} or eccentricities \cite{B76,L75}. Arikati and Maheshwari~\cite{AM1996realizing} provide an efficient technique to realize degree sequences in the PRAM model, and in~\cite{ACCPSS22}, the authors explored graph realization from a distributed setting. However, both of these works assumed a fault-free setting. Here we extend the model to a faulty setting. In~\cite{ACCPSS22}, Augustine et al. discussed distributed graph realizations on a path (both implicit and explicit). However, in their work, a node is only required to learn its neighbors in the final realization, in our work, the nodes are aware of the entire graph.

In the area of P2P overlays, there is a great deal of existing literature. In particular, a large amount of research exists to create overlays that provide structure and stability. This is best captured by overlays such as Chord~\cite{SML03}, CAN~\cite{RFHKS01}, and Skip Graphs~\cite{AS07}. Overlays have also been specifically designed to tolerate faults~\cite{AS02,AS18}. For a more detailed survey on P2P overlays and their  properties, we refer interested readers to the following excellent surveys~\cite{M15,LCPSL05}.

Our network is modeled using the Congested Clique model. The congested clique model, first introduced by Lokter et al.,~\cite{LSPP05} has been well studied, in both the faulty and non-faulty settings~\cite{patt2011round,dolev2012tri}. Problems such as agreement and leader election have also been well studied in this model \cite{ACDNP0S19,AMP18,GK10, KM021}. To the best of our knowledge, this is the first time graph realization problems have been studied in the faulty setting of Congested Clique.

While the results for the Congested Clique are interesting in and of itself, in order to make our work more applicable to P2P settings, we also explore how to solve the graph realization problem in the NCC~\cite{AGGHSKL19}. In the NCC, unlike the CC, a node is allowed to send/receive at most $O(\log n)$ messages in a round, this makes gathering information in a faulty-setting slightly more challenging.

\section{Preliminary: The Sequential Havel-Hakimi Algorithm for Graph Realization}\label{sec:shh}
We will now briefly introduce the sequential Graph Realization problems that inspired our distributed version of the same. Graph realization problems are fairly simple in characterization. The basic premise is as follows:  given a particular degree sequence $D=(d_1, d_2 \ldots d_n)$, can we create a graph $G$ whose degree sequence is precisely $D$? The most well known characterization is given independently by Havel~\cite{H55} and Hakimi~\cite{Ha62}, which  can be stated concisely as follows. 

\begin{theorem}[Based on~\cite{H55} and \cite{Ha62}] \label{thm:hh} A non-increasing sequence $D=(d_1,d_2,...,d_n)$ is graphic (i.e., graph is realizable) if and only if the sequence $D'=(d'_2,...,d'_n)$ is graphic,
where $d'_j =d_j -1$, for $j\in [2,d_1 +1]$, and $d'_j =d_j$, for $j\in [d_1 +2,n]$
\end{theorem}

This characterization directly implies a $O\big(\sum_{i=1}^n d_i \big)$ time (in terms of number of edges) sequential algorithm, known as the Havel-Hakimi algorithm, for constructing a realizing graph $G=(V,E)$  where $V=\{v_1,...v_n\}$ and $d(v_i)=d_i$, 
or deciding that no such graph exists. The algorithm works as follows.
Initialize $G=(V,E)$ to be an empty graph on $V$.  For $i=1$ to $n$, in step $i$ do the following:

\begin{enumerate}\label{alg:seqhh}
	\item Sort the remaining degree sequence in non-increasing order  ($d_i\geq d_{i-1}\geq \ldots d_{n}$).
	\item Remove $d_i$ from $D$, and set $d_j=d_j-1$ for all $j \in[i+1,d_i+i+1]$.
	\item Set the neighborhood of the node $v_i$ to be $\{v_{i+1},v_{i+2},...v_{i+1+d_i}\}$.
\end{enumerate} 

If, at any step, $D$ contains a negative entry, the sequence is not realizable.

%% file: without_knowing_f.tex
\section{Fault-Tolerant Graph Realization}\label{sec: without knowing f}
In this section, we present an efficient solution for graph realization with faults. Recall that  we are given a $n$-node Congested Clique, in which at most $f<n$ nodes may crash arbitrarily at any time. Also, a vector of degree sequence $(d_1, d_2, \dots, d_n)$ is given as an input such that each $d_i$ is only known to one node in the clique. Our goal is to construct an overlay graph of size at least $n-f$ if realizable; otherwise output unrealizable. We present an algorithm that guarantees, despite $f< n$ faulty nodes, that (i) all the (non-crashed) nodes learn and recreate a degree sequence whose size is at least $n-f$, and (ii) this degree sequence is the same for all the nodes as a requirement of the graph realization (see Definition~\ref{def : with fault}). This allows the non-crashed nodes to locally realize the overlay graph with the help of Havel-and-Hakimi's algorithm, described in Section~\ref{sec:shh}. We note that the algorithm does not require any knowledge of $f$. Our algorithm crucially uses only a few number of nodes to be involved in propagating the information about crashed nodes to the other nodes in the network-- which helps to minimize the message complexity and round complexity of the algorithm.

 At a high level, the algorithm runs in two phases. In the first phase, which consists of only two rounds, each node sends the message containing its ID and input-degree twice (in two consecutive rounds). Based on the frequency of the received message, i.e., zero or one or two times, the receiver node considers the sender node's status as faulty or non-faulty. Then each node locally creates an initial faulty list and a degree sequence (known only to itself). In the second phase, through sharing the information about faulty nodes present in the faulty list, the nodes rectify the degree sequence and create a final degree sequence $D'$, which is guaranteed to be the same for all the non-crashed nodes. The non-crashed nodes then realize the overlay graph using $D'$ as the degree sequence via the Havel-and-Hakimi algorithm.    

Let us recall a few basic assumptions. For simplicity, we assume the $KT_1$ version of the Congested Clique model, in which all the nodes know the IDs of the nodes in the clique and the corresponding link or port connecting to the IDs. Thus, in the $KT_1$ version, a node can sort all the nodes in the clique according to their distinct IDs. W.l.o.g, let us assume the IDs are $U=\{u_1,\ldots u_n\}$ in the sorted order, i.e., $u_i < u_j$ for $i<j$. A node knows its position in the sorted order. Notice that a node $u_i$ can track whether it has received any messages from a node $u_j$. However, the algorithm also works in the $KT_0$ version of the Congested Clique model where only the IDs are known, but the corresponding links are unknown (i.e., a node doesn't know which neighbor has which ID).

We will now explain the algorithm in detail. \onlyLong{a complete pseudocode is given in Algorithm~\ref{alg: without_f}.} The first phase (which we call the {\em initialization phase}) consists of two rounds. In both the rounds, every node broadcasts their input degree value to reach all the nodes in the clique. After the two rounds, each node $u_i$ creates a \emph{faulty-list} $F_{u_i}$ and a {\em degree sequence} $D'_{u_i}$ as follows. $F_{u_i}$ consists of $n$ cells corresponding to all the nodes $u_j\in U$ (in the sorted order). For any $u_j$ $ (j\neq i)$, if $u_i$ hears from $u_j$ in both the rounds, $u_i$ includes $u_j$'s degree value $d(u_j)$ in its final degree sequence $D'_{u_i}$ (and correspondingly $u_j$'s entry in the faulty-list is empty, i.e., $F_{u_i}(u_j)=null$). If $u_i$ hears from $u_j$ only in the first round (i.e., $u_j$ crashed during the initialization phase), then $u_j$ is included as a \fault ~in its corresponding entry. If $u_i$ did not hear from $u_j$ in both the rounds (i.e., $u_j$ crashed in the first round itself), then $u_j$ is additionally marked as a \smite ~in $u_i$'s \emph{faulty-list}. At the end of the algorithm, we show that $D'_{u_i} = D'_{u_j}$ for all $i, j$, which represents the final \textit{non-faulty-list} $D'$. 


In the second phase, nodes update their faulty-list locally. Nodes are divided into two groups based on the two states-- {\emph{active}} state and {\emph{listening}} state. A node in the active state transmits messages, 
whereas a node in the listening state only receives messages. Nodes may update their faulty-list according to the information received from the active nodes. Initially, all the nodes start in the listening state. A node $u_i$ in listening state continues to be in that state as long as there is another active node in the network. Let $u_j$ be the ID of the last active node (if there is none, let $j=1$). A node $u_i$ changes its status from listening state to active state if and only if $u_i$ has not received any message from the last $3(i-j)$ rounds where $j$ is the index of last heard node, i.e., $u_j$. Then $u_i$ becomes active in the ${(3(i-j)+r)}^{th}$ round; $r$ is the round when $u_i$ heard from $u_j$.


After the initialization phase (i.e., from the $3^{rd}$ round), the node with the minimum ID (or minimum index) becomes the active node, say the node $u_1$ (if present). If $F_{u_1}$ is non-empty, then $u_1$ sends the $\langle ID, d(ID)\rangle$ from the first non-empty cell in $F_{u_1}$ twice (in two consecutive rounds). If there is no $d(ID)$ then $u_1$ sends only ID. This situation may arise when an ID crashed such that $u_1$ doesn't receive $d(ID)$ in initialization phase (in first two rounds). $u_1$ continues sending $\langle ID, d(ID)\rangle$ from its $F_{u_1}$ one by one from minimum ID to maximum ID and for each message $\langle ID, d(ID)\rangle$ it sends twice in consecutive rounds. When $u_1$ finishes sending all the non-null entries in $F_{u_1}$, it switches to the exit state (which we will explain shortly). If at any point in time, $u_1$ crashed then $u_2$ will become the next active node after $3(2-1) = 3$ rounds.  
The exit state for any node $u_j$ consists of two tasks, (i) move all the $F_{u_j}$ (if any) into $D'_{u_j}$. (ii) Send out “all-okay” message to all other nodes and exit the algorithm. Sending or receiving “all-okay” message conveys that all of the known faulty-nodes have been addressed. All other nodes who hear the message (those who are in the listening state) “all-okay” also enter the exit state. 

Let us now discuss the update process of nodes in active and listening states in detail. Suppose $u_i$ is a node which is currently in its active state and $F_{u_i}$ is non-empty. Then $u_i$ sends the first non-null entry, say $s$, twice in consecutive rounds. If $s$ was a \smite ~then $u_i$ permanently removes $s$ from the $F_{u_i}$. If $s$ was a \fault, then $u_i$ moves $d(s)$ to $D'_{u_i}$. If $F_{u_i}$ is empty, then $u_i$ enters the exit state.


If instead the node $u_j$ happened to be in its listening state, then let $i$ be the index of the last node heard by the node $u_j$. If $u_j$ has not heard from any node during the last $3(j-i)$ rounds, then $u_j$ sets its state to active. Since each node needs to send the message twice, we maintain a gap of one extra round (total three) to prevent two or more nodes from being active at the same time. But if $u_j$ is currently receiving messages from an active node $u_i$ then it does the following. Let $s$ be the ID it heard from $u_i$. $u_j$ now updates its faulty-list $F_{u_j}$ based on how many times $u_j$ heard about $s$ from the active node $u_i$.

If $u_j$ heard about $s$ twice, and $s$ was a \smite ~then $u_j$ permanently removes $s$ from its faulty-list $F_{u_j}$. However, if $s$ was a \fault ~then $u_j$ moves $s$ to $D'_{u_j}$ (if not in $D'_{u_j}$) and  permanently removes $F_{u_j}(s)$ (if any). Notice that $u_j$ might have received $s$ earlier two times, in that case $F_{u_j}(s)$ will be null and there will be a corresponding entry in the $D'_{u_j}$. If $u_j$ heard about $s$ only once, then if $s$ had been a \smite ~(or respectively a \fault) ~then $F_{u_j}(s)$ is marked as a \smite ~(respectively \fault). All non-null entries in $F_{u_j}$ below $s$'s index are included in $D'_{u_j}$ if the entries correspond to faulty-nodes, otherwise they are set to null. 

Throughout the algorithm, inclusion of the degree in $D'$ is permanent. Therefore, the corresponding node entries in \textit{faulty-list} remain null. At last, all the non-faulty nodes have the same view of \textit{non-faulty-list} i.e, $D'$. Therefore, the graph can be realized locally by the Havel-Hakimi's algorithm (see Section~\ref{sec:shh}). \onlyShort{Let us call this algorithm as {\sc FT-Graph-Realization}. A complete pseudocode can be found in the full version in the Appendix.}

We will now show the correctness of the algorithm using the following lemmas and then analyze the time and the message complexity.  Lemma~\ref{lem:same-view}, Lemma~\ref{lem:same-below-list} and Lemma~\ref{lem:same-D} show that the final degree sequence $D'$ of all the non-crashed nodes are the same. Thus, the algorithm correctly solves the distributed graph realization with faults in the Congested Clique. Finally, Lemma~\ref{lem:time-complexity} analyzes the time complexity and Lemma~\ref{lem:message-complexity} analyzes the message complexity of the algorithm.     

\onlyLong{
\begin{figure}
\vspace{-1em}
\begin{algorithm}[H]
\caption{\sc Fault-Tolerant-Graph-Realization}\label{alg: without_f}
\begin{algorithmic}[1]
\Require {A Congested Clique of $n$ nodes $U = \{u_1,u_2,\ldots, u_n\}$. Each node $u_i$ is given a degree value $d(u_i)$ as an input.}
\Ensure{A corresponding graph realization that satisfies the conditions for distributed graph realization with faults.}
\State For the first two rounds, each node $u_i$ broadcasts $\langle u_i, d(u_i)\rangle$ to all the nodes. Each node creates the faulty-list $F_{u_i}$ and the degree sequence $D'_{u_i}$.
\State Nodes are either in active, listening, or exit state. If $u_1$ is non-faulty, then $u_1$ becomes the first active node.
\Statex \textbf{Nodes in Active State}.
        \While{$F_{u_i}$ is non empty}
            \State $u_i$ sends the $\langle ID, d(ID)\rangle$ (send $d(ID)$ degree is available otherwise indicate smite if it's a smite node) from minimum ID to maximum ID in $F_{u_i}$ twice-- in two consecutive rounds.
            \If {sent ID is a \smite} 
            \State Remove ID from the list $F_{u_i}$. 
            \ElsIf{sent ID is \fault}
            \State Move the degree with ID to $D'_{u_i}$.
            \EndIf
        \EndWhile
            \State  Switch to exit state.
\Statex \textbf{Nodes in Listening State}    
 \If {nodes receives the message \textit{"all-okay"} from any node}
        \State Switch to exit state.
    \EndIf
\State For any listening node $u_j$, let $i$ be the index of the last node $u_i$ heard by the node $u_j$. (starts with $i=1$).
\If{$u_j$ has not heard from any node in the last $3(j-i)$ rounds.} 
  \State $u_j$ switches to active state.
\EndIf  
 \If{Node $u_j$ heard from a node $u_i$} \label{lineno: 22}
 \Statex Let $s$ be the ID heard from $u_i$. $u_j$ updates its list according to whether it heard $s$ twice or once.
    \Statex \textit{Heard Twice:}
  
     \If{$s$ is a \smite}
        \State Remove $s$ from the faulty-list permanently. All non-null entries before $s$'s index are moved appropriately.
    \EndIf
    \If{$s$ is a faulty-node}
        \State Move $s$'s degree to $D'_{u_j}$.  All non-null entries below $s$'s index are moved appropriately.
        
    \EndIf   
\EndIf

    \Statex \textit{Heard Once:}
        \If{s is a smite-node}
            \State  Set $F_{u_j}(s)$ as smite node.
    \EndIf
    \If{s is a faulty-node}
        \State Set $F_{u_j}(s)$ as faulty-node.
    \EndIf    \label{lineno: 34}

\Statex \textbf{Nodes in Exit State}
\State Remaining faulty nodes in the faulty-list are moved to $D'$.
\State Send out an \textit{“all-okay”} message to all and exit the algorithm. 

\State All the non-crashed nodes have the same view of $D'$. Therefore, graph can be realized by all the non-crashed nodes using the Havel-Hakimi's algorithm. 
\end{algorithmic}
\end{algorithm}
\end{figure}
}

The \emph{view} of a node $s$ at $u$ is the classification of $s$ as a \smite ~or \fault ~at $u$. We will now show that for any node $s$, the view of $s$ is same across all the nodes at the end of the protocol.
\begin{lemma}\label{lem:same-view}
 Let $s$ be an ID for which there are conflicting views at the beginning of second phase. Then at the end of the protocol, i.e., when all nodes have reached the exit state, all nodes are guaranteed to have the same view of $s$.
\end{lemma}
\begin{proof}
We prove this in cases. Suppose $s$ is sent by some node. We look at the possible scenarios for both the heard-twice and the heard-once.

\noindent\textit{Heard Twice:} For the trivial case when a good node successfully broadcasts $s$ twice to all nodes, the statement follows immediately. For the other case when a faulty node successfully broadcasts $s$ in the first round, but crashes in the second round, there is a possibility that some nodes may have heard about $s$ twice, but the others may have not. In this case, all the nodes are still guaranteed to have heard $s$ at least once (in the first round of transmission) and thus would have the same view, as they would have all updated their views simultaneously.

\noindent\textit{Heard Once:} Let us consider the two cases. \\ 
\noindent\textit{Case~1:} Node crashed sending $s$ during the second round. This scenario is equivalent to the second scenario in the heard-twice case, and thus follows the same logic. \\
\noindent\textit{Case~2:} Node crashed sending $s$ during the first round. In this case, let $u_i$ and $u_j$ be two nodes that survived until the end, then there must exist a round in which $u_i$ (or respectively $u_j$) informed each other about their view of $s$. Then, based on the received values, they updated their views to match.
\end{proof}

\begin{lemma}\label{lem:same-below-list}
Let $s$ be an ID that was successfully transmitted twice by an active node $u_i$ during the second phase of the algorithm. Then for any pair of nodes $u_j, u_k$ $(j,k\geq i)$, all the (non-null) entries below $s$ are the same in their faulty-list. That is, for any non-null entry, $F_{u_j}(p) = F_{u_k}(q)$ for all $p, q <s$.  
\end{lemma}
\begin{proof}
Suppose not. Let there be a node $r$ ($r<s$) such that for two nodes $u_j$, $u_k$, $F_{u_j}(r)$ reads as  \smite ~while $F_{u_k}(r)$ reads as \fault. If $r$ had crashed during the second round of phase 1, this immediately contradicts our claim as all nodes can only have $r$ as either $null$ or \fault. In case it crashed during the first round, consider the following. Since $r$'s cell in $u_i$ is null (otherwise $r's$ value would have been transmitted first by $u_i$), $u_i$ must have heard  $r$ successfully twice at a previous round of either phase. Which implies that all other nodes would have heard about $r$ at least once, hence would share the same view of $r$. 
\end{proof}



\begin{lemma}\label{lem:same-D}
If a node $u_i$ decides to exclude an entry from its \textit{faulty-list} then all other nodes $u_j$ ($j\neq i)$ will also eventually exclude that entry. Conversely, if $u_i$ moves an entry from its fault-list to $D_{u_i}$, then eventually all other nodes $u_j$ will also move that entry to their $D_{u_j}$.  
\end{lemma}

\begin{proof}
If a node $s$ is excluded from $u_i$'s \textit{faulty-list} then $u_i$ must have heard $s$ as a \smite ~twice, which indicates that all other nodes must have received $s$ at least once  (as a \smite). So all the nodes exclude $s$ from their faulty-list. The same logic applies in case of moving an entry to $D_{u_i}$.  
\end{proof}

\begin{lemma}\label{lem:time-complexity}
The time complexity of the \onlyShort{{\sc FT-Graph-Realization} algorithm}\onlyLong{Algorithm \ref{alg: without_f}} is $O(f)$ rounds.
\end{lemma}
\begin{proof}
The first phase takes exactly two rounds. For the finalization of $D'$, any non-faulty node would require at most $f$ messages in $O(f)$ rounds. In case of faults, there can be  at most $f$ node-crashes, which can introduce a delay of at most $f$ rounds. During phase 2, a node may take at most $O(f)$ rounds to inform the network of the entries in its fault-list (including delay due to faults). Therefore, round complexity is $O(f)$.
\end{proof}

\begin{lemma}\label{lem:message-complexity}
The message complexity of the \onlyShort{{\sc FT-Graph-Realization} algorithm} \onlyLong{Algorithm \ref{alg: without_f}} is $O(n^2)$, where each message is at most $O(\log n)$ bits in size.
\end{lemma}
\begin{proof}
The first phase of the algorithm uses $O(n^2)$ messages, since all the nodes broadcast in two rounds. In the second phase, each faulty node's ID is broadcast only twice by a single (non-faulty) node. Since there are at most $f$ faulty nodes, $O(nf)$ messages incur. Therefore, total message complexity is $O(n^2)$, since $f<n$. Since both the value of degree and the ID can be encapsulated using $O(\log n)$ bits, messages are at most $O(\log n)$ bits in length.
\end{proof}

Thus, we get the following main result of fault-tolerant graph realization. 
\begin{theorem}
Consider an $n$-node Congested Clique, where $f<n$ nodes may crash arbitrarily at any time. Given an $n$-length graphic sequence $D = (d_1, d_2, \dots, d_n)$ as an input such that each $d_i$ is only known to one node in the clique, there exists an algorithm \onlyShort{({\sc FT-Graph-Realization})} which solves the fault-tolerant graph realization problem in $O(f)$ rounds and using $O(n^2)$ messages. 
\end{theorem}

\onlyLong{
\begin{remark}[Extension to $KT_0$ Model]\label{rem:KT0}
The same message and round complexity is achievable in the $KT_0$ model, in which each node knows the IDs of all the nodes in the clique, but doesn't know which port is connecting to which node-ID. Algorithm~\ref{alg: without_f} can be easily modified to work in this $KT_0$ model. During the first two rounds (i.e., in the initialization phase), each node keeps track of the links or ports through which it receives the messages. If it receives a message $\langle u_j, d(u_j) \rangle$ through some port, that port must be connecting to node-ID $u_j$. At the same time, a node can identify the faulty nodes' ID from which it didn't receive any messages. If not received in the first round itself, the ID is considered as a smite node. Once we have the initial faulty-list and the degree sequence, then we run the second phase of the Algorithm \ref{alg: without_f}. 
\end{remark}
}

\begin{table*}[t]
\centering
\scriptsize
\begin{tabular}{|p{2.75cm}|p{9.25cm}|}
\hline
\multicolumn{2}{|c|} {Terminology at a Glance for a node $u$}\\
\hline
 Terminology   & Definition \\
\hline
 &\\
    Smite-node*    & Node $v \neq u$ is classified as \smite ~if $u$ did not receive $v$'s degree in the first two rounds. \\
     &\\
      Smite-node (Phase 2)  & Node $v \neq u$ is classified as \smite  ~if $u$ receives a message from any node (even once) which classifies $v$ as a \smite.\\
    Faulty-node   &  Node $v \neq u$ is classified as a \fault ~if $u$ receives  $v$'s degree only once. \\
     &\\
    Faulty-list ($F_{u}$) &  List of known faulty IDs  (and their corresponding degrees if present) at $u$. \\
   
    &\\
    Listening State & A node is in the listening state if it is waiting for its turn (to send entries from its faulty-list) or to receive an “all-okay” message.\\
    &\\
    Active State        & A node is in the active state if it is transmitting the entries from its faulty-list.\\
    &\\
    “all-okay” &  Reception of “all-okay” at $u$ acts as a signal for $u$ to terminate the algorithm.\\
    &\\
    Degree Sequence ($D_u'$)& $D_u'$ keeps track of all the degrees heard from other nodes in the network. At termination, it contains the final degree sequence used for graph realization at $u$.\\
    &\\
\hline

\end{tabular}
    \caption{Terminology and their definition used throughout the algorithms in the paper. * represent the terminology's definition for Phase 1.}
\label{tbl: terminology}
\end{table*}

\input{lower_bound}

%% file: lower_bound.tex
\subsection{Lower Bound}\label{sec: lower_bound}
In the graph realization with faults problem, the nodes are required to learn the degrees assigned to the other nodes in the network and recreate a degree sequence which must be the same for all the nodes. If two nodes have a different view of the final degree sequence $D'$, then their output would be different. It essentially reduces to a {\em consensus} problem where all the nodes agree on the degrees in $D'$. In the consensus problem, all nodes start with some input value and the nodes are required to agree on a common value (among all the input values to the nodes). In the presence of faulty nodes, all non-faulty nodes must agree on a common value. Therefore, any $t$-round solution of the graph realization with faults problem can be used to solve the consensus problem in $O(t)$ rounds. Now, a lower bound of $f+1$ on the number of rounds required for reaching consensus in the presence of $f$ crash failures is a well-known result (see Chapter 5: Fault-Tolerant Consensus of Attiya-Welch's book \cite{AW04}). Thus, this $f+1$ lower bound on the round complexity also applies to the graph realization with faults problem. Hence, we get the following result. 

\begin{theorem}\label{thm: lower_dgrf}
Any algorithm that solves the distributed graph realization with faults in an $n$-node Congested Clique with $f$ crash failures requires $\Omega(f)$ rounds in some admissible execution. 
\end{theorem}

We now argue that graph realization with faults, where nodes construct the entire overlay graph, requires $\Omega(n^2)$ messages in the Congest model \cite{Pelege2000}. Consider a graph realization algorithm $\mathcal{A}$ that constructs the entire overlay graph in a $n$-node Congested Clique. The algorithm $\mathcal{A}$ must correctly output the overlay graph for any inputs.  

First consider the fault-free case, i.e., no faulty nodes in the Congested Clique. Since every node constructs the entire overlay graph, the $n$ degree-values (for all the nodes) need to be propagated to all the nodes. The size of a degree can be $O(\log n)$ bits. In the Congest model, the size of each message is $O(\log n)$ bits. Hence, a node can send at most a constant number of degree-values in one message packet. 
Now, it requires $n-1$ messages to send one degree-value from a node to all the nodes. Thus, it requires at least $n(n-1) = O(n^2)$ messages to propagate $n$ degree-values to all the nodes.

Consider the faulty case. The following situation may occur in some execution of $\mathcal{A}$. A faulty node may crash in a round by sending $O(n)$ messages in that round. Further, any message of a crashed node that has not reached to all the nodes may need to be rectified to make it consistent throughout the network. This requires that the message may need to be propagated to all the nodes. Hence, we require at least $O(n^2)$ messages in the worst case. Thus, we get the following result.  
\begin{theorem} \label{alg: lower_bound_message}
In the {\em Congest} model, any graph realization algorithm, in which nodes construct the entire overlay graph, in a $n$-node network (with or without faults) requires $\Omega(n^2)$ messages in some admissible execution.
\end{theorem} 

Therefore, the above two theorems prove that the \onlyShort{{\sc FT-Graph-Realization} algorithm} \onlyLong{ Algorithm \ref{alg: without_f}} is tight simultaneously with respect to the time and the message complexity.

%% file: ncc-extension.tex
\onlyLong{

\section{Fault-Tolerant Graph Realization in the NCC Model}\label{sec: NCC model}

In this section we extend the above fault-tolerant algorithm to the Node Capacitated Clique (a.k.a NCC) model. To the best of our knowledge, distributed versions of graph realization problem (without faults) were first studied by the authors in~\cite{ACCPSS22}. In the original work, the authors attempted to solve the distributed degree sequence problem in two versions of the NCC model introduced by the authors in~\cite{AGGHSKL19}, namely the $NCC_0$ and the $NCC_1$. In this section, we present how we may extend our ideas for solving the graph realizations with faults problem in the Congested Clique to $NCC_1$.

Briefly, the $NCC_1$ is exactly like the $KT_1$ version of the Congested Clique (CC) model with one clear difference, which is a constraint on the number of messages a node is allowed to send/receive in a round.  In the $NCC_1$, a node is allowed to send or receive at most $O(\log n)$ messages in a round, unlike the CC model, in which we don't have a bound on the number of messages. When a node receives more than $O(\log n)$ messages, it chooses to drop the excess. Note that the model immediately implies a $n/\log n$ lower bound in terms of time complexity for our version of the graph realization, as each node needs to learn $n-1$ degrees in a clique. Our solution takes $O(nf/\log n)$ rounds in the $NCC_1$ model, but it is optimal in the number of messages.

We will now present how we modify Algorithm~\ref{alg: without_f} for the $NCC_1$. The key idea is to change how each node sends its degree to every other node in the network. We leverage the idea of parallelism and cyclic permutation so that each node can in one round (i) inform $O(\log n)$ other nodes and (ii) receive the degree (or faulty IDs as the case may be) from at most $O(\log n)$ other nodes. 

The steps above gives us the following straightforward idea, divide the nodes into $n/\log n$ groups $g_1,g_2\ldots,g_{\frac{n}{\log n}}$ such that each group has no more than $O(\log n)$ nodes. This allows all nodes in a group $g_i$ to send their degree to all nodes in the group $g_j$ while satisfying the message constraints present in $NCC_1$.

By taking advantage of the parallelism present in the setting (and working with different permutations in each round) we can guarantee that all nodes learn the degrees in the network in $O(n/\log n)$ rounds. This gives the idea of global broadcast (Algorithm~\ref{alg:global}).

\begin{algorithm}[H]
\caption{\sc Global-Broadcast}\label{alg:global}
\begin{algorithmic}[1]
    \For{$r=0$ to $n/\log n-1$}
        \For{$j\in[1,\frac{n}{\log n}]$ in parallel}
            \State $dest=(r+j)\mod \frac{n}{\log n}$ 
            \State Nodes in group $g_j$ send their degrees to all the nodes in group $g_{dest}$.
        \EndFor
    \EndFor
    \end{algorithmic}
\end{algorithm}

Clearly, the procedure in Algorithm~\ref{alg:global} ensures that $n$ nodes may send $n$ messages each over a period of $n/\log n$ rounds. We may use global broadcast to guarantee that up to $n$ non-faulty nodes may send their degrees to the network in $n/\log n$ rounds, but no node receives more than $O(\log n)$ degrees. We use a similar approach when it comes to sending out “all-okay”, the difference being we need to ensure that no two nodes send “all-okay” to the same group at once.

Algorithm~\ref{alg: global_update} ensures that all nodes may receive “all-okay” while making sure that two active nodes are not sending the “all-okay” message to the same group of nodes at the same time. We can now modify Algorithm~\ref{alg: without_f} for the NCC. It follows the same steps, the key difference is in the fact that when an active node is sending entries from its faulty-list, it requires $O(n/\log n)$ rounds to inform all the nodes in the network. Since there can be $f$ entries in the worst case scenario, this means that it can take $O(nf/\log n)$ rounds to ensure that all nodes have the same degree sequence. The process for local update of entries in the faulty-list remains the same as in Algorithm~\ref{alg: without_f}
\begin{algorithm}[H]
\caption{\sc Global-Update} \label{alg: global_update}
\begin{algorithmic}[1]
    \State \textbf {Global Update:} This protocol is executed at $u_i$ when it receives an all-okay message.
    \For{$j=i$ to $n/\log n$}
      \State $u_i$ sends all-okay to all nodes in $g_j$.
     \EndFor
     \For{$j=1$ to $j=i-1$}
      \State $u_i$ sends "all-okay" to all nodes in $g_j$.
     \EndFor
    \end{algorithmic}
\end{algorithm}

Now we argue that the Algorithm~\ref{alg: without_f_NCC} (below) has the same guarantees for the view of $D'$ as in Algorithm~\ref{alg: without_f}. In fact, we show that the final degree sequence $D'$ of all the non-crashed nodes are the same (using the similar arguments as in Lemma~\ref{lem:same-view}, Lemma~\ref{lem:same-below-list}, and Lemma~\ref{lem:same-D}). 
\begin{lemma}
Algorithm~\ref{alg: without_f_NCC} guarantees that all nodes have the same view of the degree sequence $D'$ at the end of the algorithm.
\end{lemma}
\begin{proof}
Suppose not. Then there exists a node $s$ for which nodes $u_i$ and $u_j$ have different views. If both the nodes survived until the end, there must exist a round where either $u_i$ or $u_j$ would have informed the other of $s$'s status, thus ensuring that they both have the same view of $s$. Also, the arguments of Lemma~\ref{lem:same-below-list}, Lemma~\ref{lem:same-D} are still applicable in the NCC case,  ensuring that no degree is excluded from some nodes of the network, while being included in the final degree sequence of the other nodes. Hence, the $D'$, created at the end of the Algorithm~\ref{alg: without_f_NCC}, is the same across all nodes. 
\end{proof}

Note that a key difference in Algorithm~\ref{alg: without_f_NCC} to the one designed for the Congested Clique model is that a single loop to inform all nodes of a fault takes $O(n/\log n)$ rounds. Since there are $f$ faults, we have the following theorem:
\begin{theorem}
Algorithm~\ref{alg: without_f_NCC} solves graph realization with faults in the Node Capacitated Clique model in $O(nf/\log n)$ rounds and uses $O(n^2)$ messages. 
\end{theorem}


\begin{algorithm}[H]
\caption{\sc Graph-Realization-with-Faults-in-NCC}\label{alg: without_f_NCC}
\begin{algorithmic}[1]
\Require {A Congested Clique of $n$ nodes $U = \{u_1,u_2,\ldots, u_n\}$. Each node $u_i$ is given a degree value $d(u_i)$ as an input.}
\Ensure{A corresponding graph realization that satisfies the conditions for distributed graph realization with faults.}

\State Using the global broadcast procedure in Algorithm~\ref{alg:global}, each node $u_i$ broadcasts $\langle u_i, d(u_i)\rangle$ to all the nodes.
\State Each node creates the faulty-list $F_{u_i}$ and the degree sequence $D'_{u_i}$.
\State Nodes are either in active, listening, or exit state. If $u_1$ is non-faulty, then $u_1$ becomes the first active node.

\Statex \textbf{Nodes in Active State}.
    \If {$u_i$ is an active node}
        \If{$F_{u_i}$ is non empty}
            \State The following for loop is repeated twice. 
            \For{$j\in[1,\frac{n}{\log n}]$ in parallel}
            \State $u_i$ sends the $\langle ID, d(ID) \rangle$ (send $d(ID)$ if available otherwise it's a smite node) from minimum ID to maximum ID in $F_{u_i}$ to $g_j$. If the list becomes empty, then switch to exit state.
            \EndFor
            \If {sent ID is a \smite} 
            \State Remove ID from its faulty-list. 
            \ElsIf{ sent ID is \fault }
            \State Remove ID from faulty-list. Move ID and ID's degree to $D'$.
            \EndIf
            
        \Else
            \State Enter exit state.
      \EndIf
    \EndIf
\Statex \textbf{Nodes in Listening State}    
 \If {received the message \textit{all-okay} from any node}
        \State Switch to exit state.
    \EndIf
\State For any listening node $u_j$, let $i$ be the index of the last node heard by the node $u_j$.    
\If{$u_j$ has not heard from any node during the last $3(j-i)\frac{n}{\log n}$ rounds.} 
\State Node switches to active state.
\Else

\State Procedure follows the exact same steps in Line~\ref{lineno: 22} to Line~\ref{lineno: 34} in Algorithm~\ref{alg: without_f}.
 \EndIf
    
  
        
    \Statex \textbf{Nodes in the Exit State} 
\State Nodes that are marked as a \fault ~in the faulty-list are moved to the final degree sequence $D'$.
\State Send out an \textit{all-okay} message to the nodes in the network in groups of size $O(\log n)$ (see Algorithm~\ref{alg: global_update} for the exact detail) and exit the protocol.

\State All the non-faulty nodes have the same view of \textit{non-faulty-list}. Therefore, graph is realized locally by the Havel-Hakimi algorithm. 
\end{algorithmic}
\end{algorithm}
}

\onlyShort{
\section{Fault-Tolerant Graph Realization in the NCC Model}\label{sec: NCC model}

In this section we extend the above fault-tolerant algorithm to the Node Capacitated Clique (a.k.a NCC) model. To the best of our knowledge, distributed versions of graph realization problem (without faults) were first studied by the authors in~\cite{ACCPSS22}. In the original work, the authors attempted to solve the distributed degree sequence problem in two versions of the NCC model introduced by the authors in~\cite{AGGHSKL19}, namely the $NCC_0$ and the $NCC_1$. In this section, we present how we may extend our ideas for solving the graph realizations with faults problem in the Congested Clique to $NCC_1$.

Briefly, the $NCC_1$ is exactly like the $KT_1$ version of the Congested Clique (CC) model with one clear difference, which is a constraint on the number of messages a node is allowed to send/receive in a round.  In the $NCC_1$, a node is allowed to send or receive at most $O(\log n)$ messages in a round, unlike the CC model, in which we don't have a bound on the number of messages. When a node receives more than $O(\log n)$ messages, it chooses to drop the excess. Note that the model immediately implies a $n/\log n$ lower bound in terms of time complexity for our version of the graph realization, as each node needs to learn $n-1$ degrees in a clique. Our solution takes $O(nf/\log n)$ rounds in the $NCC_1$ model, but it is optimal in the number of messages. 

The key idea is to change how each node sends its degree to every other node in the network. We leverage the idea of parallelism and cyclic permutation so that each node can in one round (i) inform $O(\log n)$ other nodes and (ii) receive the degree (or faulty IDs as the case may be) from at most $O(\log n)$ other nodes. Thus, we divide the nodes into $n/\log n$ groups $g_1,g_2\ldots,g_{\frac{n}{\log n}}$ such that each group has no more than $O(\log n)$ nodes. This allows all nodes in a group $g_i$ to send their degree to all nodes in the group $g_j$ while satisfying the message constraints present in $NCC_1$. By taking advantage of the parallelism present in the setting (and working with different permutations in each round) we can guarantee that all nodes learn the degrees in the network in $O(n/\log n)$ round. The detail algorithm and the analysis is placed to Appendix due to space limitation. Theorem~\ref{thm:ncc-main} presents the main result of this section.  

\begin{theorem}\label{thm:ncc-main}
There exists a fault-tolerant algorithm that solves graph realization in the Node Capacitated Clique model in $O(nf/\log n)$ rounds and uses $O(n^2)$ messages. 
\end{theorem}

}


%% file: conclusion.tex

\section{Conclusion and Future Work}\label{sec: conclusion}

In this paper, we studied the graph realization problem in the Congested Clique with faults and provided efficient algorithms for realizing overlays for a given degree sequence. Our algorithm is simultaneously optimal in both the round and the message complexity. Further, we also show how the algorithm may be adapted to a setting, in which nodes are allowed to send (and receive) a limited number of messages per round. Given the relevance of graph realization techniques in overlay construction and the presence of faulty nodes in the peer-to-peer networks, we believe there can be several interesting questions to explore in the future. Such as (1) Is it possible to remove the assumption of known IDs, that is, would it be possible to achieve optimal round and message complexity if the IDs of the clique nodes are not known? (2) This paper and the previous paper on distributed graph realization \cite{ACCPSS22} consider a clique network. It would be interesting to study the problem beyond clique network. (3) Finally, it would be interesting to study the graph realization problem in the presence of Byzantine faults. Since a Byzantine node is malicious and can send wrong information, the graph realization problem needs to be defined carefully.  